\documentclass[letter,twocolumn]{jpsj3}
\usepackage{color}

\def\runauthor{D. Aoki \textit{et al.}}

\title{Pressure Evolution of the Magnetic Field induced Ferromagnetic Fluctuation through the Pseudo-Metamagnetism of CeRu$_2$Si$_2$}

\author{
Dai~{\sc Aoki}$^1$\thanks{E-mail address: dai.aoki@cea.fr}, 
Carley~{\sc Paulsen}$^2$, 
Tatsuma~D.~{\sc Matsuda}$^{1,3}$, 
Liam~{\sc Malone}$^4$, 
Georg~{\sc Knebel}$^1$,
Pierre~{\sc Haen}$^2$, 
Pascal~{\sc Lejay}$^2$, 
Rikio~{\sc Settai}$^5$, 
Yoshichika~{\sc \=Onuki}$^5$, and 
Jacques~{\sc Flouquet}$^1$\thanks{E-mail address: jacques.flouquet@cea.fr}
}

\inst{%
$^1$INAC, SPSMS, CEA Grenoble, 38054 Grenoble, France\\
$^2$Institut N\'{e}el, CNRS/UJF Grenoble, 38042 Grenoble, France. \\
$^3$ASRC, JAEA, Tokai, Ibaraki 319-1195, Japan\\ 
$^4$LNCMI, UPR 3228 (CNRS-INSA-UJF-UPS), 31400 Toulouse, France\\
$^5$Graduate School of Science, Osaka University, Toyonaka, Osaka 560-0043, Japan
}

\abst{
Resistivity measurements performed under pressure in the paramagnetic ground state of CeRu$_2$Si$_2$ are reported.
They demonstrate that 
the relative change of effective mass through the pseudo metamagnetic transition 
is invariant under pressure.
The results are compared with the first order metamagnetic transition due to the antiferromagnetism
of Ce$_{0.9}$La$_{0.1}$Ru$_2$Si$_2$
which corresponds to the ``negative'' pressure of CeRu$_2$Si$_2$ by volume expansion.
Finally, we describe the link between the spin-depairing of quasiparticles on CeRu$_2$Si$_2$ 
and that of Cooper pairs on the unconventional heavy fermion superconductor CeCoIn$_5$.
}

\kword{pseudo-metamagnetism, effective mass, CeRu$_2$Si$_2$, CeCoIn$_5$}

\begin{document}
\maketitle

\begin{figure}[tbh]
\begin{center}
\includegraphics[width=0.8\hsize,clip]{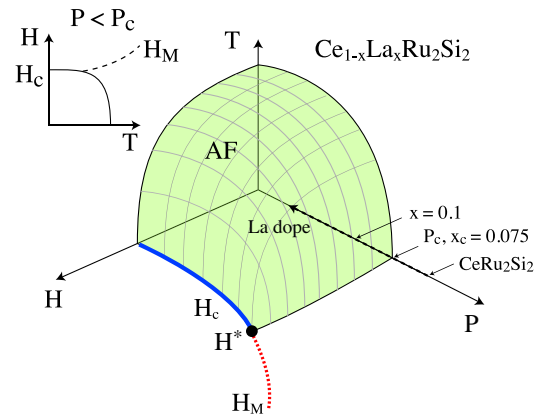}
\end{center}
\caption{(Color online) Schematic temperature--pressure (La doping)--field phase diagram of Ce$_{1-x}$La$_x$Ru$_2$Si$_2$. A blue thick line denoted by $H_{\rm c}$ is the first order metamagnetic transition. A red dotted line denoted by $H_{\rm M}$ is the crossover, which is called pseudo-metamagnetic transition in the text. $H^\ast$ corresponds to the metamagnetic quantum critical end point (QCEP).~\protect\cite{Flo06_review,Wei10}
The inset shows how the lines $H_{\rm M}$ and $H_{\rm c}$ will meet for $P<P_{\rm c}$.~\protect\cite{Fis91}}
\label{fig:CeRu2Si2_TPH}
\end{figure}

A large variety of experiments have been reported on the heavy fermion compound CeRu$_2$Si$_2$ using macroscopic and microscopic probes,
as it is a good candidate for the study of the evolution of the electronic properties as a function of the magnetic polarization,
which in this system can be tuned by magnetic field and pressure.~\cite{Flo06_review,Flo10} 
Furthermore, it is a rare example where the Fermi Surface has been fully determined for its low magnetic field ($H$) paramagnetic (PM) ground state.~\cite{AokiH95} 
Thermodynamic macroscopic measurements have established that a strong enhancement of the effective mass ($m^\ast$) occurs at the pseudo-metamagnetic field 
$H_{\rm M}\sim 7.8\,{\rm T}$. 
In CeRu$_2$Si$_2$, a transition from the PM ground state to an antiferromagnetic (AF) ground state 
could be achieved by applying a negative pressure, i. e. by expanding the volume. 
This was first realized by substituting La for Ce~\cite{Fis91}. 
The critical pressure $P_{\rm c}$ corresponds to a critical concentration $x_{\rm c} = 0.075$ in Ce$_{1-x}$La$_x$Ru$_2$Si$_2$ alloys. 
In the latter, for $x>x_{\rm c}$ ($p<p_{\rm c}$) 
a real first order metamagnetic transition from the AF ground state to a field induced polarized state 
is observed with increasing field 
while the pseudo-metamagnetic transition is observed for $x<x_{\rm c}$. 
A schematic temperature--pressure (La doping)--field phase diagram is shown in Fig.~\ref{fig:CeRu2Si2_TPH}. 
The dotted line $H_{\rm M}$ corresponds to the continuation of the real metamagnetic line $H_{\rm c}$. 
For $x = x_{\rm c}$, the field $H^\ast$ ($\sim 4T$) which separates these two lines 
can be considered as quantum critical end point (QCEP) as already asserted in Refs.~\citen{Flo06_review,Wei10}.

Here we report new measurements of resistivity under pressure 
in order to investigate the pressure response of the $m^\ast$ enhancement.
To precisely compare the results with the first order metamagnetism 
which will occur below $P_{\rm c}$ (i.e. $x > x_{\rm c}$ in Ce$_{1-x}$La$_x$Ru$_2$Si$_2$), 
we have carried out a new set of magnetization experiments on Ce$_{0.9}$La$_{0.1}$Ru$_2$Si$_2$ ($T_{\rm N} \sim 5\,{\rm K}$) 
down to $100\,{\rm mK}$ using the same procedure 
that was realized two decades ago in CeRu$_2$Si$_2$.~\cite{Pau90} 
To complete the studies under magnetic fields through $x_{\rm c}$, 
specific heat ($C$) measurements were performed for $x=0.1$ and $0.075$.  
Finally, we will emphasize the link between the individual quasiparticle spin depairing at $H_{\rm M}$ 
and the depairing of Cooper pairs in CeCoIn$_5$
following recent experiments realized on this exotic unconventional superconductor
near the superconducting upper critical field $H_{\rm c2}$~\cite{Tay02,Pau_pub}. 
 
High quality single crystals of CeRu$_2$Si$_2$ and Ce$_{1-x}$La$_x$Ru$_2$Si$_2$ ($x=0.1$ and $0.075$)
were grown by the Czochralski method.
Single crystals were oriented by Laue photographs and 
samples were obtained from them by spark cutting.
The resistivity measurements on CeRu$_2$Si$_2$ were made by the four-probe AC method 
with the electrical current along the $a$-axis of the tetragonal crystal structure.
The residual resistivity ratio ($\rho_{\rm RT}/\rho_0$) in CeRu$_2$Si$_2$ was 100 indicating the high quality
in the present samples.
The measurements under pressure were carried out using a hybrid-type piston cylinder cell
up to $5.4\,{\rm kbar}$.
The pressure was determined by the superconducting transition temperature of Pb.
The measurements below $16\,{\rm T}$ were done with a superconducting magnet at temperatures down to $100\,{\rm mK}$,
while the measurements at higher fields up to $28\,{\rm T}$ were made using a resistive magnet at LNCMI-Grenoble
and with temperatures down to $0.46\,{\rm K}$.
The magnetization measurements for Ce$_{0.9}$La$_{0.1}$Ru$_2$Si$_2$ were carried out
using a miniature dilution refrigerator and a low-temperature SQUID magnetometer 
at temperatures down to $100\,{\rm mK}$ and at fields up to $8\,{\rm T}$.
The specific heat measurements for Ce$_{1-x}$La$_x$Ru$_2$Si$_2$ ($x=0.1$ and $0.075$)
were made using the relaxation method at temperatures down to $0.45\,{\rm K}$ and at fields up to $9\,{\rm T}$.
For all measurements, the magnetic field was applied along the tetragonal $c$-axis.

\begin{figure}[tbh]
\begin{center}
\includegraphics[width=1 \hsize,clip]{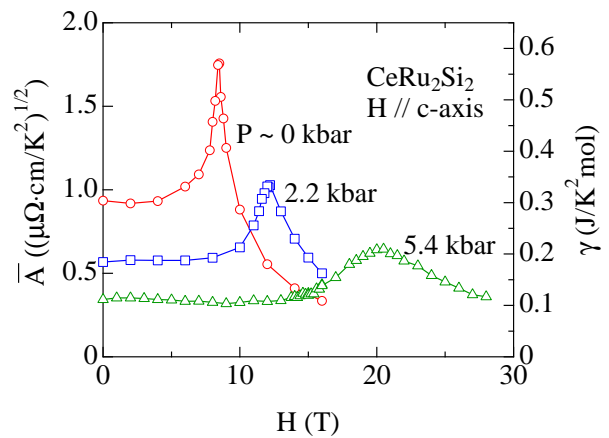}
\end{center}
\caption{(Color online) Field dependence of the $A$ coefficient of $T^2$ term of resistivity in the form of $\sqrt{A}$ vs $H$ at three different pressures 0, 2.2 and $5.4\,{\rm kbar}$ in CeRu$_2$Si$_2$. We assumed here the validity of $\sqrt{A} \propto m^\ast \propto \gamma$. A right axis was scaled at zero field at ambient pressure.}
\label{fig:A_coef}
\end{figure}

Figure~\ref{fig:A_coef} represents the field dependence of $\sqrt{A}\sim m^*$, where $A$ is the coefficient of the $T^2$ term of the resistivity based on the Fermi liquid law
and $m^\ast$ is the effective mass.
At zero pressure it was verified that $A$ scales with the Sommerfeld coefficient $\gamma$ ($\propto m^\ast$) of the specific heat,
as described by Kadowaki-Woods relation.~\cite{Kam96}. 
The validity of this relation is based on the fact that
local fluctuations will be accompanied with the development of intersite interactions 
which in turn depend on the wave vector of those interactions, AF as well as ferromagnetic ones.

\begin{figure}[tbh]
\begin{center}
\includegraphics[width=0.9 \hsize,clip]{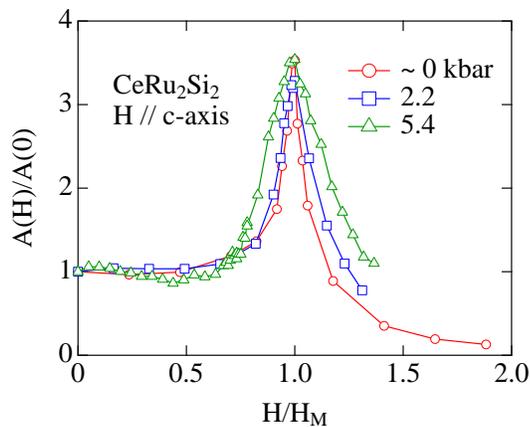}
\end{center}
\caption{(Color online) Scaling of $A(H)/A(0)$ in $H/H_{\rm M}$ at 0, 2.2, and $5.4\,{\rm kbar}$ in CeRu$_2$Si$_2$.}
\label{fig:A_coef_scale}
\end{figure}

As seen in Fig.\ref{fig:A_coef}, there is a strong suppression of $A$ at zero field with pressure, and in addition, the magnitude of
$A$ at $H_{\rm M}$ decreases with pressure. 
The remarkable result, shown in Fig.~\ref{fig:A_coef_scale}, is that 
excellent scaling is observed in a plot of $A(H)/A(0)$ versus $H/H_{\rm M}$. 

\begin{figure}[tbh]
\begin{center}
\includegraphics[width=0.8 \hsize,clip]{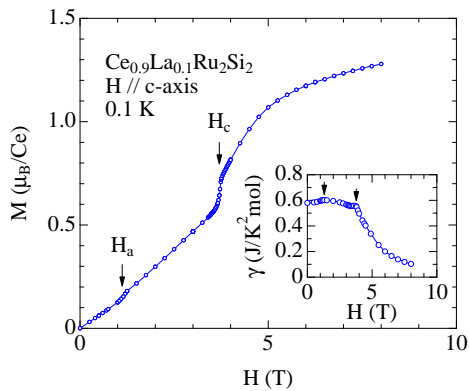}
\end{center}
\caption{(Color online) Magnetization of Ce$_{0.9}$La$_{0.1}$Ru$_2$Si$_2$ at $0.1\,{\rm K}$ with the magnetization jump at $H_{\rm a}\sim 1.2\,{\rm T}$ and 
$H_{\rm c}=3.8\,{\rm T}$. The inset shows the field variation of $\gamma$ deduced from the temperature dependence of $M(T)$ at different fields from the thermodynamic Maxwell relation.}
\label{fig:mag_gamma}
\end{figure}

The inset of Fig.~\ref{fig:mag_gamma} shows the field dependence of $\gamma$ for the AF system Ce$_{0.9}$La$_{0.1}$Ru$_2$Si$_2$ at ambient pressure, 
which was obtained from careful studies of the temperature dependence of the magnetization at various fields using the thermodynamic Maxwell relation, namely $\partial M/\partial T = \partial S /\partial B$,
where $S$ is the entropy.~\cite{Pau90}
This AF system undergoes two metamagnetic transitions, 
first at $H_{\rm a}\sim 1.2\,{\rm T}$ between two AF states,
and then at $H_{\rm c}=3.8\,{\rm T}$ when the system goes from the AF to the PM state.~\cite{Flo06_review}
There is a small maximum at $H_{\rm a}$ which is broad, in agreement with the width of the magnetization jump. 
Between $H_{\rm a}$ and $H_{\rm c}$, $\gamma$ is weakly field dependent. 
Above $H_{\rm c}$, there is a sharp decrease of $\gamma$, 
quite similar to that observed in CeRu$_2$Si$_2$ above the pseudo-metamagnetic transition. 
The dominant mechanism around $H_{\rm a}$ is due to AF interactions.
Neutron diffraction experiments have shown that, at $H_{\rm a}$, the main change is the switch from a low field wave vector $(0.31,0,0)$
to a high field wave vector $(1/3, 1/3, 0)$~\cite{Mig91}.
On the other hand, ferromagnetic interactions dominates above $H_{\rm c}$. 
The weak variation of $\gamma$ between $H_{\rm a}$ and $H_{\rm c}$ results from the balance 
between these two competing mechanisms. 
The similarity of the decrease of $\gamma$ above $H_{\rm c}$ in the alloys and above $H_{\rm M}$ in CeRu$_2$Si$_2$ 
comes from the development in this compound of low energy ferromagnetic fluctuations observed by neutron experiments.~\cite{Sat04} 

\begin{figure}[tbh]
\begin{center}
\includegraphics[width=0.8 \hsize,clip]{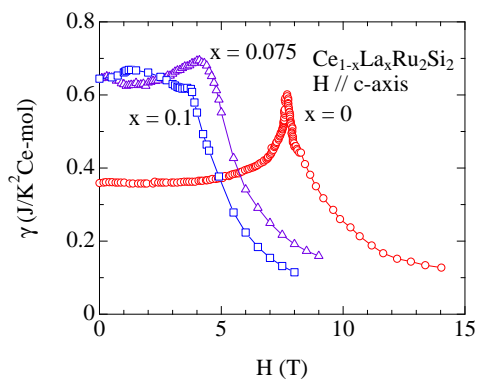}
\end{center}
\caption{(Color online) Field variation of $\gamma$ extrapolated at $T= 0\,{\rm K}$ for  Ce$_{0.9}$La$_{0.1}$Ru$_2$Si$_2$ and CeRu$_2$Si$_2$ and measured at 0.45 K for 
Ce$_{0.925}$La$_{0.075}$Ru$_2$Si$_2$ assuming $C/T=\gamma$.
The data for CeRu$_2$Si$_2$ were cited from ref.~\citen{Flo10}.}
\label{fig:Hdep_gamma}
\end{figure}

Figure~\ref{fig:Hdep_gamma} summarizes the field dependence of $\gamma(H)$ at ambient pressure for different concentrations 
$x$: $x=0.1$ ($x>x_c$), $x=0.075$ ($x=x_c$) and $x=0$. 
Our new data are in excellent agreement with previous data~\cite{Flo06_review,Fis91,AokiY98}
and are more precise.
In particular, for $x=0.1$, the measured temperature range
has been extended to $0.1\,{\rm K}$ via low temperature magnetization measurements.
For $x=0.075$, special attention has been given to realize small field steps around $H_{\rm M}$ ($\sim H^\ast$). 
These results confirm that for the CeRu$_2$Si$_2$ family, a critical value of $\gamma_{\rm c}\sim 600\, {\rm mJ/K^2 mol}$ is achieved,  either by pressure through $P_{\rm c}$ or using magnetic field through $H^\ast$.
The fact that $\gamma(H_{\rm M})$ for the pure CeRu$_2$Si$_2$ is very close to $\gamma(H^\ast)$ for $P=P_{\rm c}$
is due to the strong thermal expansion of the lattice at $H_{\rm M}$
which drives the system very close to the QCEP.~\cite{Flo06_review}
The occurrence of the sharp anomaly of $\gamma$ in the field dependence which are correlated with those of $\gamma$ at $P_{\rm c}$
is the mark of the magnetic critical end point at $H^\ast$.
The magnetic critical end point is linked to the Ising character of the Ce magnetic coupling.~\cite{Flo10}
In the case of a Heisenberg system (such as CeNi$_2$Ge$_2$) just above $P_{\rm c}$
a strong initial decrease of $\gamma$ with field will occur
when $H_{\rm c}$  collapses at $P_{\rm c}$,
and thus the magnetic field drives the system to the paramagnetic state.~\cite{Geg03}
There is a real decoupling between AF instability and spin depairing due to the field,
which appears in CeNi$_2$Ge$_2$ at $H \sim 40\,{\rm T}$.

The resistivity measurements have clarified the pressure and field dependence of $m^\ast$,
as the Fermi liquid regime is here achieved at low temperatures.
In a previous study,
resistivity experiments under pressure were realized only above $1.5\,{\rm K}$.~\cite{Mig89} 
With our new set of data down to $100\,{\rm mK}$, the description becomes clearer.
Increasing the pressure,
the $A$ coefficient at zero field is strongly suppressed 
due to the huge value of the electronic Gr\"{u}neisen parameter $\Omega_{\rm e}$
as the system moves away from the AF singularity at $P_{\rm c}$. 
However, the field dependence of $m^\ast$ associated with the field-induced ferromagnetic (FM) fluctuations,
which is built from the AF pseudogap, still remains.
Applying a magnetic field at different pressures,  
the identical shape of a field-dependent $m^\ast$ is illustrated by the scaling of $A(H,P)/A(0,P)$ as shown in Fig.~\ref{fig:A_coef_scale}. 
This result is in excellent agreement with the theoretical picture.~\cite{Satoh01}

Inelastic neutron scattering experiments in magnetic fields have revealed that applying a magnetic field  basically leads to a suppression of the  
AF correlations at $H_{\rm M}$ with a characteristic energy of $1.6\,{\rm meV}$ 
while just in the vicinity of $H_{\rm M}$ low energy ferromagnetic fluctuations (energy near 0.4 meV at $H_{\rm M}$) emerge.~\cite{Flo04,Sat04}
Another manifestation of the key role of FM fluctuations is the
observation that  pseudo-metamagnetism and metamagnetism coincide at a critical value of 
$M$ ($\equiv M_c\sim 0.6 \mu_{\rm B}$) in excellent agreement with 
i) the equality between the electronic Gr\"{u}neisen  parameter $\Omega_{\rm e}\sim 200$ at $H=0$ and 
the Gr\"{u}neisen parameter of the pseudo-metamagnetic field $\Omega_{H_{\rm M}}\sim 200$ and 
ii) the self-consistency between the field dependence of the 
linear $T$ coefficient of the thermal expansion~\cite{Pue88} and the specific heat through a two parameter entropy ($S$) scaling with $T/T^\ast$, $H/H_{\rm M}$ 
where $\Omega_{\rm e}$ is described by $\Omega_{\rm e}= -\partial \log T^\ast / \partial \log V$. 
The emergence of a critical value of the magnetization $M_{\rm c}$ implies that a spin depairing will occur at this value.
Quantum oscillation experiments have shown that it is associated with a drastic Fermi surface evolution.~\cite{AokiH95}. 
Let us point out that the effects of pseudo-metamagnetism as well as metamagnetism (see ref.~\citen{Kna10,Hol95}) is felt over larger magnetic 
field window than in the response of the magnetization. 
The amplification of the thermal fluctuations is directly linked to the large strength of the electronic  Gr\"{u}neisen parameter 
while the magnetic response at constant pressure is mainly governed by the magnetostriction effect.~\cite{Flo06_review,Flo10,Kna10} 

It is interesting to compare the  properties of the CeRu$_2$Si$_2$ family (where the  metamagnetic phenomena in strongly correlated electron systems is  well established) to the case of of CeCoIn$_5$
where metamagnetism is observed at $H_{\rm c2}$ at low temperatures~\cite{Tay02,Pau_pub}.
The common point is that, due to the strength of $\gamma$ (i.e. the weakness of the characteristic energy),
a significant value of the magnetization is induced at a
quite low field in comparison to other materials. 
Furthermore, the possibility of spin decoupling either of the individual quasi-particle or of the Cooper pair 
will induce the balance between AF and FM fluctuations 
and their associated Cooper pairing at pressures close to $P_{\rm c}$. 
For superconductivity (SC), the field-interplay between SC and Pauli paramagnetism will favor a non zero momentum of the Cooper pair as predicted by 
Fulde, Ferrell, Larkin and Ovchinnikov four decades ago (FFLO state).~\cite{Ful64,Lar65} 
In CeCoIn$_5$, for $H \parallel a$ in the basal plane, 
a new high-magnetic and low-temperature phase (Q-phase) has been reported~\cite{Ken10}.
For $H \parallel c$, it is believed that 
a singularity exists at $H_{\rm c2}$ without the appearance of a Q-phase.  
The image is that the magnetic field adds a FM fluctuation channel to the AF ones 
and this allows a mixture between the singlet and triplet component due to the concomitant creation of the Q-phase.
Two models have been proposed in this framework taking into account the occurrence of a FFLO phase.~\cite{Yan09,Ape10}
Another proposal is to assume that the paramagnetic depairing modifies the $d$-wave excitation and leads to the creation the Q-phase.~\cite{Ike10}. 
The similarity of CeCoIn$_5$ with CeRu$_2$Si$_2$ is that the metamagnetic-like transition is correlated with a large enhancement of $\gamma$ 
just at $H_{\rm c2}$ for CeCoIn$_5$ instead of $H_{\rm M}$ for CeRu$_2$Si$_2$.
However $H_{\rm c2}$ will not be a magnetic quantum critical singularity. 

\begin{figure}[tbh]
\begin{center}
\includegraphics[width=1 \hsize,clip]{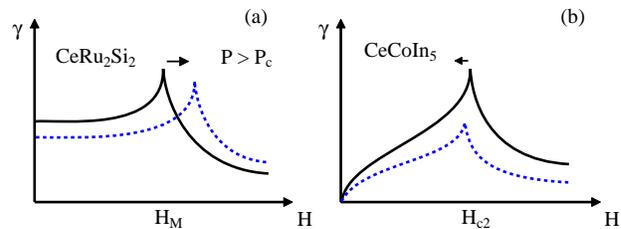}
\end{center}
\caption{(Color online) Schematic pressure response of $\gamma (H)$ through the spin depairing of the psedo-metamagnetism of CeRu$_2$Si$_2$ at $H_{\rm M}$ and the paramagnetic 
depairing at the superconducting upper critical field $H_{\rm c2}(0)$ of CeCoIn$_5$. Blue dashed lines correspond to the plot under pressure.}
\label{fig:CeRu2Si2_CeCoIn5}
\end{figure}

To illustrate this point, a schematic view is shown in Fig.~\ref{fig:CeRu2Si2_CeCoIn5} for the predicted field variation of $\gamma$ in CeRu$_2$Si$_2$ and in CeCoIn$_5$ for two different pressures. 
Under pressure, a maximum of $\gamma$ will persist presumably at $H_{\rm c2}$ up to $1.3\,{\rm GPa}$
above which no long range ordering will occur in CeCoIn$_5$.~\cite{Mic06,Kne08}
In CeRu$_2$Si$_2$, in agreement with the schematic plot in Fig.~\ref{fig:CeRu2Si2_CeCoIn5}(a),
a huge variation of the Gr\"{u}neisen parameter has been pointed out
even with a sign change at $H_{\rm M}$~\cite{Flo06_review}.
In CeCoIn$_5$, no change of sign of $\Omega_{\rm e}(H)$ under fields has been observed at $H_{\rm c2}$~\cite{Zau10}
as well as a weak field dependence of $\Omega_{\rm e}(H)$.
An amazing case would be if triplet SC can occur near $H_{\rm M}$
where ferromagnetic fluctuations might be enhanced
in a heavy fermion compound. 
Of course in the absence of a Q-phase no mixture will happen between singlet and triplet components. 

In summary,
pressure experiments on CeRu$_2$Si$_2$ demonstrate the field evolution of FM fluctuations 
through the spin depairing which occurs at the pseudo-metamagnetic transition.
The FM component plays an important role for the field enhancement of $m^\ast$ at $H_{\rm M}$, 
as it is built from the AF ``camel''-shaped singularity in the density of state.
Complementary results on the antiferromagnetic side via La substitution establish:
i) the quasi-equality between $\gamma (P_{\rm c})$ and $\gamma (H^\ast)$
in this Ising heavy fermion system and
ii) the quasi-invariance of $\gamma(H)$ between $H_{\rm a}$ and $H_{\rm c}$, which is close to the critical value of $\gamma_{\rm c}$.
This indicates that a remarkable high field low temperature AF phase is stabilized close to QCEP.
The description of tricriticality~\cite{Mis09} may deserve new insights taken into account the feedback on Fermi surface.
The mechanism in CeCoIn$_5$ at $H_{\rm c2}$ also reflects the interplay between 
strong AF and FM fluctuations correlated with specific channels of SC pairing and gap topology.
We have restricted our comparisons to the debated case of CeCoIn$_5$.
However, the CeRu$_2$Si$_2$ example is a key reference for the field effects in the heavy fermion compounds.
A possibility of the cascade of spin depairing of the different bands 
of the Fermi surface can happen in URu$_2$Si$_2$.~\cite{Shi09}
and the transverse field instability in ferromagnets such as UCoGe~\cite{Aok11_ICHE} 
looks quite similar to the antiferromagnetic one.

\section*{Acknowledgements}
This work was supported by the EC program ``Transnational Access'' (EuromagNET II),
ERC starting grant (NewHeavyFermion), French ANR project (CORMAT, DELICE),
a Grant-in-Aid for Scientific Research on Specially
Promoted Research (No. 20001004), and
Scientific Research on Innovative Areas ``Heavy Electrons''
(20102002) from the Ministry of Education, Culture, Sports,
Science and Technology, Japan.


%

\end{document}